# An overview of continuous and discrete phasor analysis of binned or time-gated periodic decays

X Michalet[1]

Department of Chemistry & Biochemistry
607 Charles E. Young Drive E., Los Angeles, CA 90095, USA

## ABSTRACT

Time-resolved analysis of periodically excited luminescence decays by the phasor method in the presence of time-gating or binning is revisited. Analytical expressions for discrete configurations of square gates are derived and the locus of the phasors of such modified periodic single-exponential decays is compared to the canonical universal semicircle. The effects of IRF offset, decay truncation and gate shape are also discussed. Finally, modified expressions for the phase and modulus lifetimes are provided for some simple cases. A discussion of a modified phasor calibration approach is presented.

## 1. INTRODUCTION

The analysis of luminescence decays remains an active research topic [1-3]. Experimentally, information on the temporal dependence of fluorescence, phosphorescence and other types of luminescence decay can be obtained by either frequency-modulated or pulsed excitation. Periodic pulsed excitation-based approaches result in time-resolved recordings of the periodic emission intensity, which are fitted by a decay model, typically multi-exponential, but not necessarily [4-6]. An alternative, and in many ways simpler approach, phasor analysis [7-9], has recently emerged as a popular method.

Signals recorded with TCSPC (time-correlated single-photon counting) hardware can, after proper correction, be determined with exquisite precision and considered quasi-continuous. In this case, the phasor $z$ of such decays can be defined as an integral and possess the remarkable property that the phasor of the *recorded decay* is identical to that of the (unknown) *emitted signal* up to a rotation and/or dilation in the complex plane [10, 11]. The situation arising when sparse (instead of quasi-continuous) sampling or lower photon arrival time resolution (encountered for instance in time-gated data or when data is binned with lower resolution) are encountered, is a bit different and theoretical results about it are limited [12-16]. Time-gating scheme involving *overlapping gates*, or *non-adjacent gates* bring their own set of complications.

More generally, partial coverage of the laser period (which correspond to 'truncated' decays) or situations where the excitation pulse is offset with respect to time 0 in the recording time window, which are common experimentally, further affect phasor calculation, whether one deals with quasi-continuous (TCSPC) or discrete (time-gated or integrated) signals. As new detectors with non-standard gating schemes are introduced, and because results obtained with various modalities need to be quantitatively compared [17], studying modifications to phasor analysis and the interpretation of its results in these non-canonical situations appears timely.

This article presents results for the phasor of *periodic single-exponential decays* (PSEDs), but extension to linear combinations of PSEDs or more general decays is straightforward and can be found in [18].

The article is organized as follows: Section 2 introduces basic concepts and definitions regarding gated (as well as ungated) decays encountered in luminescence lifetime experiment involving periodic excitation. Section 3 discussed phasor analysis concepts, including properties of the phasor of convolution products, in particular phasor of decays with finite sampling (or 'discrete' phasor). It also reviews basic properties of the loci of phasors of PSEDs ('Single-Exponential Phasor Loci' curves or SEPL). In Section 4, the effect of a *decay offset*, *decay truncation* on the phasor of PSEDs are studied. We also briefly address the influence of a non-square *gate shape* on previous results. Section 5 then examines extensions of the standard phase and modulus lifetime definitions for some of the cases discussed in the previous sections.

[1] michalet@chem.ucla.edu

Section 6 examines modifications to the concept of *phasor calibration* in the different situations described previously. Finally, the conclusion summarizes the concepts introduced in the article.

This article assumes some conceptual familiarity with phasor analysis, which has been discussed in recent reviews (for instance ref. [19, 20]). It is an abridged version of a recently article, to which the interested readers are referred for details [18].

## 2. TIME-GATED PERIODIC DECAYS

Conventionally, luminescence decays are described as the response to a single excitation pulse, $x_0(t)$. Starting from the theoretical decay resulting from a hypothetical Dirac excitation pulse, $F_0(t)$, the emitted luminescence decay $\varepsilon_0(t)$ resulting from the single pulse $x_0(t)$ reads as the convolution product of the former two:

$$\varepsilon_0(t) = \int_{-\infty}^{+\infty} x_0(u) F_0(t-u) du = x_0 * F_0(t) \tag{1}$$

Experimentally, however, the sample is subjected to an infinite series of periodic excitations, which can be described by the *T*-periodic summation:

$$x_{0,T}(t) = \sum_{k=-\infty}^{n(t)} x_0(t-kT) \tag{2}$$

where $n(t) = \lfloor t/T \rfloor$ is the truncation index ensuring that excitations posterior to the current time point are ignored and $T$ is the excitation period. Correspondingly, the emitted signal is given by the *T*-periodic summation of Eq. (1):

$$\varepsilon_{0,T}(t) = \sum_{k=-\infty}^{n(t)} \varepsilon_0(t-kT) = x_{0,T} * F_0(t) = x_{0,T} \underset{T}{*} F_{0,T}(t) \tag{3}$$

where we have introduced the *cyclic convolution product*:

$$x_{0,T} \underset{T}{*} F_{0,T}(t) = \int_0^T du\, x_{0,T}(u) F_{0,T}(t-u) \tag{4}$$

The emitted signal is then detected by a detector and processed by some electronic hardware, whose characteristics can be both combined into an electronic response function $E(t)$ (convolution product of the detector and processing response functions), equal to the signal that would be recorded in response to a hypothetical incident Dirac signal. The resulting signal is:

$$S_T(t) = E * \varepsilon_{0,T}(t) = E_T \underset{T}{*} \varepsilon_{0,T}(t) = I_T \underset{T}{*} F_{0,T}(t) \tag{5}$$

which introduces the T-periodic instrument response function (IRF) of the whole setup, $I_T(t)$:

$$I_T(t) = E_T \underset{T}{*} x_{0,T}(t) \tag{6}$$

Time-gating or binning can be considered as being part of the processing performed by the electronics, or when needed, treated as a separate process. In that case, defining the gate (or bin) function $\Gamma_{s,W}(t)$ as the effective detection efficiency of the instrument over the window [*s*, *s*+*W*]:

$$\Gamma_{s,W}(t) \begin{cases} = 0 & \text{if } t < s \\ \in ]0,1] & \text{if } \in [s, s+W] \\ = 0 & \text{if } t > s+W \end{cases} \tag{7}$$

the gated (or binned) signal is given by:

$$S_{T,W}(s) = \int_0^T \Gamma_{s,W}(t) S_T(t) dt = \overline{\Gamma}_W \underset{T}{*} S_T(t) \tag{8}$$

where we have introduced the mirrored gate function:

$$\overline{\Gamma}_W(s-t) = \Gamma_{s,W}(t) \tag{9}$$

Equations (1)-(9) can be used to handle most situations of interest and are applicable whether one is interested in discrete or continuous recording. Their derivation and further details can be found in ref. [18]. As an illustration, Fig. 1 shows an example of periodic single-exponential decay (PSED, $\tau = 3.8$ ns) resulting from a $T = 50$ ns periodic excitation by a single-exponential source ($\tau_0 = 0.1$ ns) and recording by a time-gated detector with a square gate width $W = 20$ ns. Similar signals can be easily simulated using the Phasor Explorer software described in ref. [18] and freely available online [21].

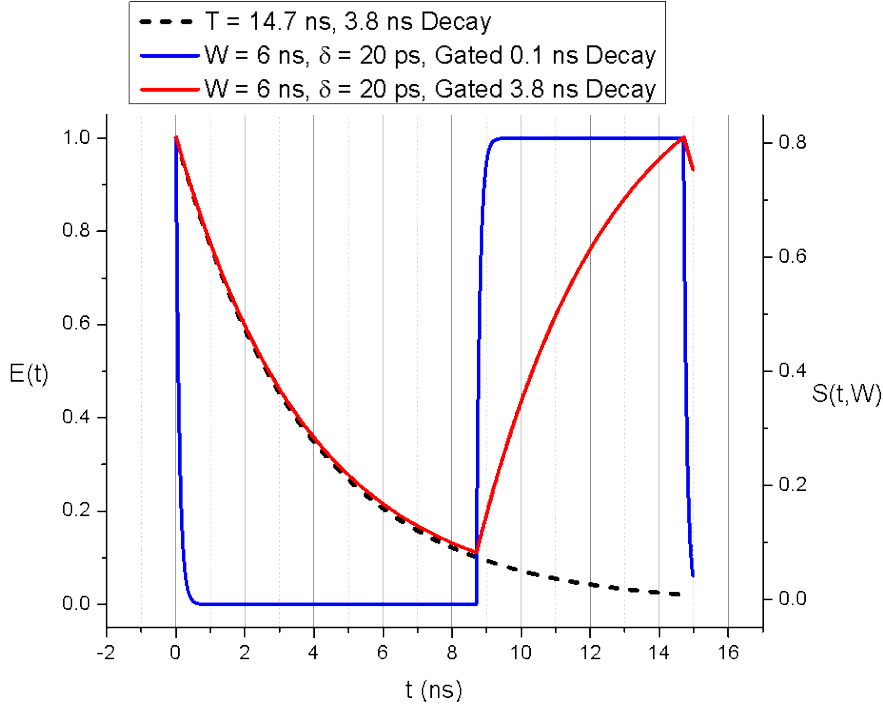

Fig. 1: Gated decays ($\tau_1 = 0.1$ ns, representing the equivalent of an IRF, and $\tau_2 = 3.8$ ns, laser period: 50 ns) with gate width $W = 25$ ns and gate separation $\delta = 500$ ps, corresponding to settings used in SwissSPAD 2 measurements performed in [16]. The time-gated decay (red) is essentially identical to the ungated decay (dashed, black), except at the end of the period where the rise time of the time-gated decay is replaced by a mirror image of the decay, of width $W = 20$ ns.

## 3. PHASOR OF PERIODIC DECAYS

### 3.1. Continuous decays

Phasor analysis has been discussed extensively in the literature [7, 20, 22, 23]. We will therefore keep its presentation to a minimum, putting particular emphasis on a formalism revealing the underlying periodic nature of the signals being analyzed.

It is convenient to combine the two components $g$ and $s$ of the phasor representation of a signal $S(t)$ into a single complex number $z = g + is$:

$$z[S](f) = \frac{\int_{-\infty}^{+\infty} dt\, S(t) e^{i 2\pi f t}}{\int_{-\infty}^{+\infty} dt\, S(t)} \tag{10}$$

where $f$ is the harmonic frequency and $i$ is the complex root of -1. Because a signal is generally defined for $t > 0$ only, the integral lower bound is effectively 0, but the above expression has the merit to connect explicitly phasor and Fourier transform. Eq. (10) can be rewritten in term of the $T$-periodic summation $S_T$ of $S$ as:

$$z[S](f) = \underset{C}{z}[S_T](f) = \frac{\int_0^T dt\, S_T(t) e^{i2\pi ft}}{\int_0^T dt\, S_T(t)} \quad (11)$$

The "cyclic" phasors (with a 'C' below the phasor symbol $z$, a notation which will drop henceforth, as it will be implicit when dealing with the phasor of periodic functions) computed at the various harmonics $f_n = n/T$, are therefore related to the Fourier series coefficients of $S_T$.

A useful property resulting from this connection is that the cyclic phasor of the convolution product of two $T$-periodic functions $f_T$ and $g_T$ is equal to the product of their cyclic phasors (the harmonic $f$ has been dropped from the notation for simplicity):

$$z\left[f_T \underset{T}{*} g_T\right] = z[f_T] z[g_T] \quad (12)$$

From Eq. (5), it follows that the phasor of a recorded decay is the product of the IRF's phasor and that of the "pure" decay, that is the decay which would be observed with a $T$-periodic Dirac comb:

$$z[S_T] = z[I_T] z[F_{0,T}] \quad (13)$$

This identity is of course at the origin of the well-known calibration procedure, to which we will return in Section 8.

### 3.2. Discrete decays

So far, we have assumed that the different functions are known at all time points $t$ within the period $T$. While this is a good approximation for decays measured with high temporal resolution TCSPC hardware, binning, or in the case of time-gated integrating detectors, acquisition duration constraints, can limit the number $N$ of samples per period that are available for analysis. In these cases, the integrals of Eq. (11) are replaced by finite sums and the definition of what we will call the *discrete (cyclic) phasor* is given by:

$$z_N[S_T](f) = \sum_{p=1}^{N} S_T(t_p) e^{i2\pi f t_p} \Big/ \sum_{p=1}^{N} S_T(t_p) \quad (14)$$

where the $N$ sampling points $t_p$ are generally equidistant, $t_{p+1} - t_p = T/N$, $p = 1...N$ and represent the start time of the bin or gate.

The main difference with the continuous definition (11) is that the discrete phasor of a convolution product is generally not equal to the product of the individual discrete phasor, with obvious consequences for the calibration process commonly used in phasor analysis. This is most obvious when looking at some practical examples, as done in the next subsections.

### 3.3. Some examples in the continuous case

One of the hallmarks of phasor analysis is the special role played by single-exponential decays. We will denote a normalized single-exponential decay with time constant $\tau$ by $\Lambda_\tau(t)$ and its $T$-periodic summation by $\Lambda_{\tau,T}(t)$. The continuous phasor of $\Lambda_\tau(t)$, equal to the continuous cyclic phasor of $\Lambda_{\tau,T}(t)$ is given by the well-known expression (in which we have dropped the mention of frequency $f$ in the notation for the phasor $z$ for simplicity):

$$z\left[\Lambda_{\tau,T}\right] = \zeta_f(\tau) = \frac{1}{1 - i2\pi f\tau} = \frac{1 + i2\pi f\tau}{1 + (2\pi f\tau)^2} \quad (15)$$

The location of their phasors ($\tau \in [0, \infty[$) is a half-circle of radius ½, centered at $z = $ ½, referred to as the universal semi-circle. Using the convolution rule (Eq. (12)) or by direct calculation, it is straightforward to compute the phasor of PSEDs convolved with a PSED IRF (time constant $\tau_0$):

$$z\left[\Lambda_{\tau_0,T} \underset{T}{*} \Lambda_{\tau,T}\right] = \zeta_f(\tau_0)\zeta_f(\tau) \tag{16}$$

Likewise, the phasor of time-gated PSEDs computed with a square gate of width $W$ is given by the following expression:

$$\begin{cases} z\left[\overline{\Pi}_{W,T} \underset{T}{*} \Lambda_{\tau,T}\right] = z\left[\overline{\Pi}_{W,T}\right]\zeta_f(\tau) = M_W e^{-i\varphi_W}\zeta_f(\tau) \\ \varphi_W = \pi f W \\ M_W = \dfrac{\sin \varphi_W}{\varphi_W} \end{cases} \tag{17}$$

where $\overline{\Pi}_{W,T}$ is the mirror-image of a square gate of width $W$.

In both cases (and in fact, in all cases), the phasors of these PSEDs are the products of a constant (complex number) by the standard functional form given by Eq. (15). In other words, the locus of the continuous phasors of PSEDs is always a half-circle, but with a radius generally different from ½ and rotated about the origin. Of course, calibration (Section 8) will bring these phasors back to the universal semi-circle.

### 3.4. Some examples in the discrete case

The simplest case of discrete phasor is encountered for "pure" PSEDs (Dirac IRF, no gating or binning) and corresponds to a continuous decay sampled at a few discrete points, a situation analogous (but not exactly identical) to that encountered with TCSPC data acquired with high timing resolution (the relevant formula for binned data, which TCSPC data is, is discussed next). A simple calculation yields:

$$\begin{cases} z_N\left[\Lambda_{\tau,T}\right] = \zeta_{f,N}(\tau) = \dfrac{1-x(t)}{1-x(t)e^{i\alpha}} \\ x(\tau) = e^{-\theta/\tau}; \alpha = 2\pi f \theta \end{cases} \tag{18}$$

This expression is clearly different from that in the continuous case (Eq. (15)) and describes an *arc of circle*, noted $\mathcal{L}_N$, where $N$ is the number of sampling points. Note that, although we dropped its mention to simplify notations, it also depends on the harmonic frequency $f$. Its center location ($g_c$, $s_c$) and radius $r$ are given by:

$$\begin{cases} g_c = \dfrac{1}{2}, \quad s_c = -\dfrac{1}{2}\tan(\alpha/2) \\ r = \dfrac{1}{2|\cos(\alpha/2)|} \end{cases} \tag{19}$$

$\mathcal{L}_N$ converges towards the universal semicircle when $N \to \infty$ as expected. By extension, we will designate the universal semicircle by $\mathcal{L}_\infty$. A few examples of these loci are shown in Fig. 2.

The next example follows in the steps of that studied in the continuous case, namely the discrete phasor of PSEDs with a PSED IRF. The calculation yields:

$$z_N\left[\Lambda_{\tau_0,T} \underset{T}{*} \Lambda_{\tau,T}\right] = e^{i\alpha} z_N\left[\Lambda_{\tau_0,T}\right] z_N\left[\Lambda_{\tau,T}\right] \tag{20}$$

where $\alpha$ has been defined in Eq. (18). Similarly to the continuous case, the discrete phasor of a PSED in the presence of a PSED IRF is equal to the discrete phasor of the "pure" PSED, $z_N\left[\Lambda_{\tau,T}\right]$, multiplied by a constant complex number: its locus is therefore a rotated, scaled arc of circle, and calibration (Section 8) will bring phasors of PSEDs back to the corresponding $\mathcal{L}_N$.

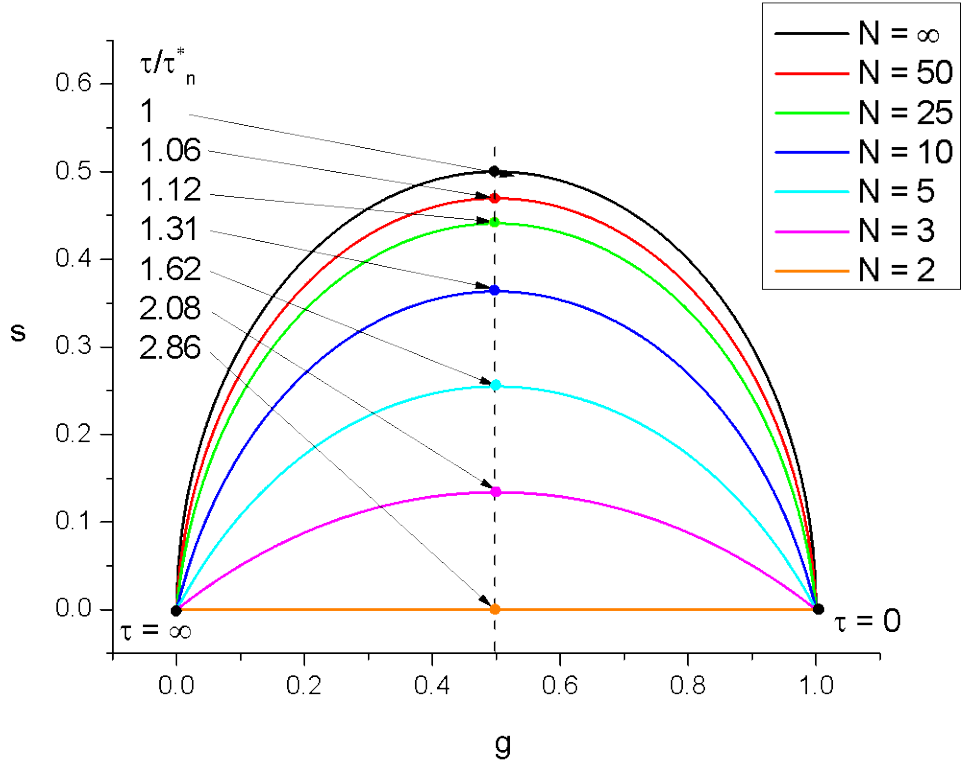

Fig. 2: Locus of discrete phasors of single-exponential decays for different choices of $N$, the number of sampling points. and $n$, the phasor harmonic. ($f = 1/T$). The extremum of these curves (indicated by a dot of the same color) is located at $g = \frac{1}{2}$ and is attained for different values of $\tau$ expressed in units of $\tau^*_n = T/2\pi n$. Notice that for $n = 1$, $N = 2$ results in a straight line.

Unfortunately, the similitude between the continuous and discrete phasors ends here. Indeed, the calculation of the discrete phasor of a square-gated PSED (with gate width $W$) yields a function whose analytical form has no simple geometrical description, except in the special cases where the gate width $W$ is proportional to the data point separation $\theta$, i.e. when $W = q\theta$, $q$ integer. In those cases, we find that:

$$z_{N[W]}\left[\Lambda_{\tau,T}\right] = z_N\left[\overline{\Pi}_{W,T} \underset{T}{*} \Lambda_{\tau,T}\right] = e^{i\alpha} z_N\left[\overline{\Pi}_{W,T}\right]\zeta_{f,N}(\tau) \tag{21}$$

where the discrete phasor of a square gate $z_N\left[\overline{\Pi}_{W,T}\right]$ has a simple expression [21]. The important fact is that this phasor is equal to a (complex) constant multiplied by the canonical discrete phasor of PSEDs, $\zeta_{f,N}(\tau)$, which shows that here again, the locus of discrete phasors of PSEDs is a rotated, scaled arc of circle.

Most interestingly, for binned data ($q = 1$), the discrete phasor given by Eq. (21) is equal to the discrete phasor of PSEDs. In other words, binning is equivalent to discrete sampling in phasor analysis.

In the general case, the situation can get quite complicated, as illustrated in Fig. 3, which shows other cases than the special ones just discussed.

## 4. THE EFFECTS OF DECAY OFFSET, DECAY TRUNCATION AND GATE SHAPE

The ideal situations described in the previous sections are in practice rarely encountered. Oftentimes, recorded decays are affected by multiple experimental constraints. For instance, it is rare that the IRF location corresponds to time point 0 (and in fact rarely desirable for multiple reasons independent from phasor analysis). In other words, decays are usually offset with respect to the time axis. This turns out to have a trivial effect on the calculated continuous phasor, but in general, a non-trivial one on discrete phasors. Another frequent departure from the ideal situation is decay truncation, i.e. a full period recording is not available because of user choice (e.g. to speed up acquisition), or because parts of the recording are affected by artefacts and need to be "cut off" to keep only the reliable parts. This also affects the calculated phasor, but can in some cases be easily compensated for by a specific choice of harmonic frequency. Finally, in the case of time-gated data, the hypothesis of a square gate might be inappropriate, and the effect of this difference on the calculated phasor deserves attention. This section will briefly examine these different topics, the reader being referred to ref. [18] for further details.

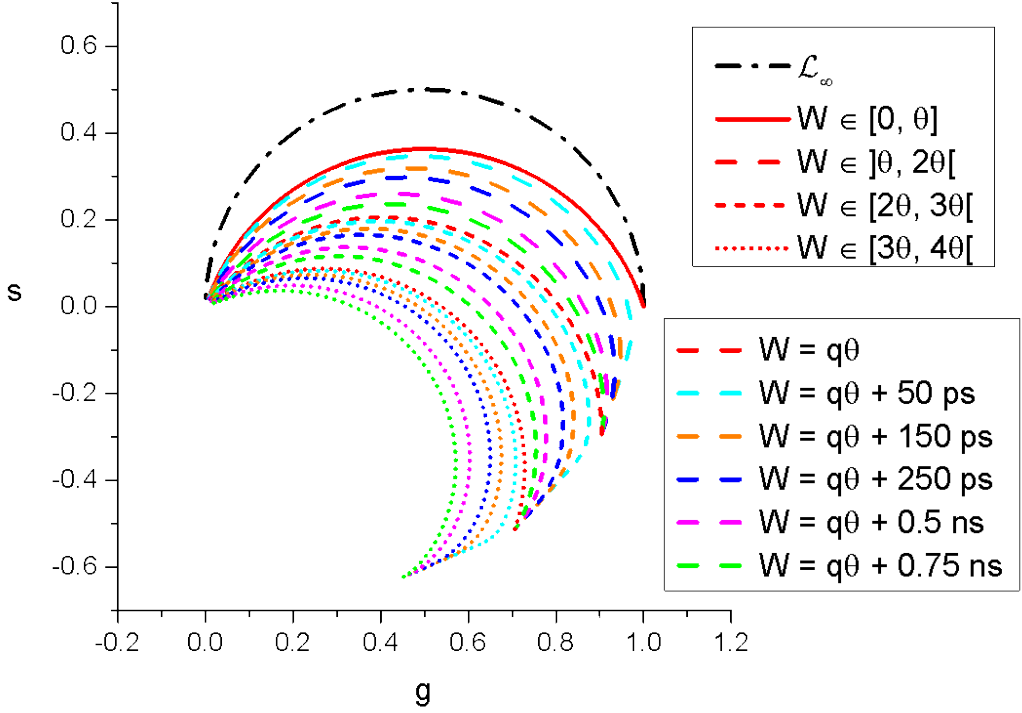

Fig. 3: Locus of discrete phasors of square-gated PSEDs. $L_{N[W]}$ for a constant number of gates $N = 10$ (and gate step $\theta = T/N$, $T = 12.5$ ns, $f = 1/T$) and varying gate width. For $W \leq \theta$, $L_{N[W]}$ is independent of W and equal to $L_N$ (solid red circular arc). For $W \in ]\theta, 2\theta[$ (long dash curves), all $L_{N[W]}$ share a common $z(0)$ and $z(\infty)$ but only $L_{N[W]}$ for $W = \theta$ (solid red curve) is a circular arc. Similarly, for $W \in [2\theta, 3\theta[$ (short dash curves) and $W \in [3\theta, 4\theta[$ (dotted curves), the different $L_{N[W]}$ share a common $z(0)$ and $z(\infty)$ in each group but only $L_{N[W]}$ for $W = 2\theta$ (short red dash curve) and for $W = 3\theta$ (red dotted curve) are circular arcs.

### 4.1. Decay offset

A $T$-periodic decay with offset $t_0$, $S_{T|t_0}(t)$, can be rewritten in terms of the offset-free $T$-periodic decay $S_T(t)$ as:

$$S_{T|t_0}(t) = S_T\big((t - t_0)[T]\big) \tag{22}$$

where the $[T]$ notation indicates the modulo operation. It is easy to verify that in the case of continuous phasors, the effect of this modification is a mere rotation by an angle $2\pi f t_0$:

$$z\big[S_{T|t_0}\big] = z[S_T] e^{i 2\pi f t_0} \tag{23}$$

In the case of discrete phasors, the situation is, as expected, subtler. In the simplest case of PSEDs (Dirac IRF), the discrete phasor is given by:

$$z_N\big[\Lambda_{\tau, T|t_0}\big] = \zeta_{f,N}(\tau) e^{i\alpha \lceil t_0/\theta \rceil} \tag{24}$$

using the definitions introduced in Eq. (18) and the notation $\lceil x \rceil$ for the `ceiling` function (the smallest integer larger than or equal to $x$). In other words, the discrete phasor of the offset decay is a rotated version of the offset-free decay, but the rotation also depends on the sample separation $\theta$ in a nonlinear manner.

As we go through the different illustrative cases studied earlier (PSED IRF, square-gated PSED), the discrete phasor of the offset version of these decays takes on more complex expressions (see ref. [18]) for details), which only simplify in a few cases. For instance, for a PSED IRF, a simple expression is obtained only when the offset is proportional to the sampling interval $\theta$:

$$t_0 = q\theta \Rightarrow z_N\left[\Lambda_{\tau_0,T|t_0} *_T \Lambda_{\tau,T}\right] = z_N\left[\Lambda_{\tau,T}\right] z_N\left[\Lambda_{\tau_0,T|t_0}\right] e^{i\alpha} \tag{25}$$

which is also a rotated version of the offset-free discrete phasor.

Likewise, in the case of a square gate of width $W$ equal to the sampling step $\theta$ (a case relevant to binning situations), and the offset is proportional to the sampling interval, the discrete phasor of PSED with offset (Dirac IRF) reads:

$$z_{N[\theta]}\left[\Lambda_{\tau,T|t_0}\right] = \zeta_{f,N}(\tau) e^{i2\pi f t_0} = z_N\left[\Lambda_{\tau,T|t_0}\right] \tag{26}$$

Once again, we find that the discrete phasor of a binned decay is identical to the discrete phasor of a decay sampled at discrete points, although *only* when the offset falls on the beginning of one of the bins.

### 4.2. Decay truncation

A truncated decay is defined over a sub-interval $[t_1, t_1+D]$ of the full period $[0, T]$. $t_1$ is the first time point in the record and $D$ is its duration. By definition, the corresponding continuous "truncated" phasor is thus:

$$\vec{z}[S_T]_{t_1,D} = \frac{\int_{t_1}^{t_1+D} dt\, S_T(t) e^{i2\pi ft}}{\int_{t_1}^{t_1+D} dt\, S_T(t)} \tag{27}$$

This expression can be computed in special cases, the simplest one being that of PSEDs with Dirac IRF, for which the continuous phasor takes a simple form:

$$\vec{z}[\Lambda_{\tau,T}]_{t_1,D} = \frac{1-e^{-D/\tau}e^{i2\pi fD}}{1-e^{-D/\tau}} e^{i2\pi ft_1} \zeta_f(\tau) \tag{28}$$

This equation doesn't describe a simple curve, unless $f$ is a multiple of $1/D$, in which case Eq. (28) takes the simple canonical form of Eq. (15), up to a rotation by an angle $2\pi ft_1$.

The situation is similar for discrete decays, which are defined at equidistant points $t_1,\ldots, t_N$ such that $0 \leq t_1 < t_N \leq T$ and $D = N\theta$. In the case of PSEDs, we obtain:

$$\vec{z}_N[\Lambda_{\tau,T}]_{t_1,D} = \frac{1-e^{-D/\tau}e^{i2\pi fD}}{1-e^{-D/\tau}} e^{i2\pi ft_1} \zeta_{f,N}(\tau) \tag{29}$$

As in the continuous case, this equation describes a simple curve only if $f$ is a multiple of $1/D$, in which case Eq. (29) takes the simple canonical form of Eq.(18), up to a rotation by an angle $2\pi ft_1$.

These results show that in some cases, the choice of a phasor harmonic frequency different from the "natural" $n/T$ series might be advantageous. Unfortunately, in cases where the IRF is not a Dirac function, or gating is used, the resulting locus of phasors of PSEDs is in general not a simple curve, although it might sometimes be close to a semicircle (continuous phasor) or an arc of circle (discrete phasor) in some cases. Direct numerical (or sometimes analytical) calculation should then be used to check its actual shape.

### 4.3. Gate shape

Although analytical results for square gates can be fairly easily obtained, as illustrated above with a few examples, experimental gate shapes can depart significantly from this ideal situation. In the continuous case, due to the convolution rule (Eq. (12)), a gate simply intervenes as a constant (complex) pre-factor in the phasor expression and thus doesn't change the nature of the locus of phasors of PSEDs, which remains a semicircle (rotated and rescaled with respect to the universal semicircle). As should not come as a surprise, following the previous results, the picture is more complex in the case of discrete decays. Fig. 4 presents a few cases of gate shapes (triangular, sawtooth and reversed sawtooth) by comparison with the square gate shape for illustration. As for the effect of truncation, it is therefore recommended to perform direct numerical or analytical calculation of the phasor of PSEDs to determine the actual shape of this locus, for instance using the Phasor Explorer software [21].

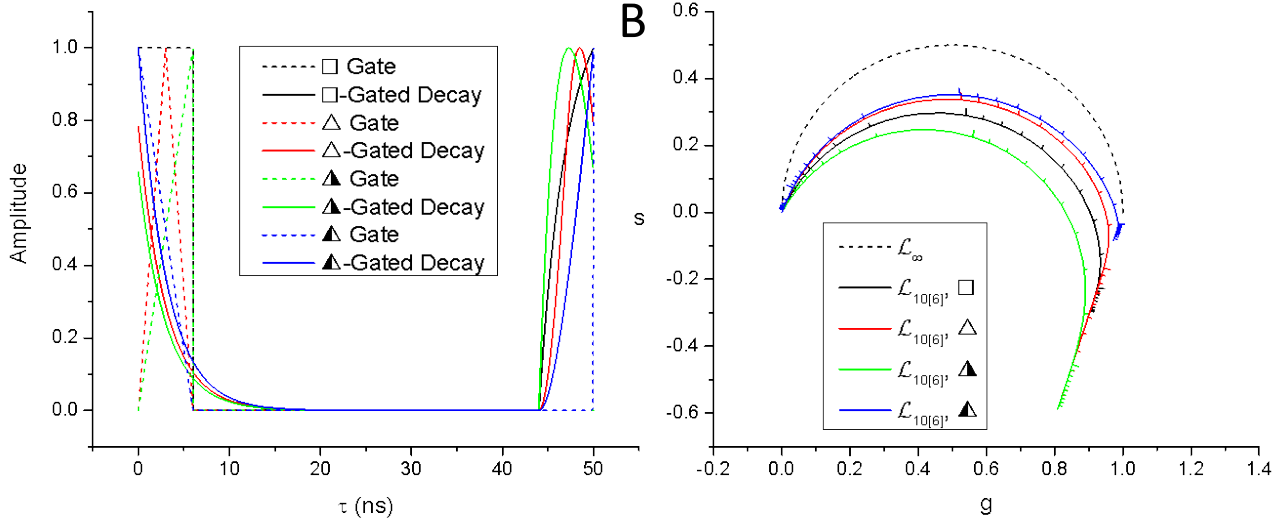

Fig. 4: Effect of gate shapes on the discrete phasors. A: 4 gates of width $W = 6$ ns (square, triangle, sawtooth and reversed sawtooth, dashed curves) are shown starting at $t = 0$ within a period of duration $T = 50$ ns ($f = 20$ MHz). The corresponding gated-decays for a PSED with $\tau = 3$ ns are shown as plain curves. Due to the different locations of the maximum of each gate, the corresponding $T$-periodic gated decays exhibit different maxima locations as well as different shapes. This effect is similar for all PSEDs and therefore results in different SEPL shown in B. B: universal semicircle ($L_\infty$, dashed curve) and the 4 SEPLs with gate width $W = 6$ ns for the same number of equidistant gate locations $N = 10$. While the SEPLs look fairly similar, noticeable differences exist. Ticks indicate the locations of PSEDs with lifetime 0.1 – 1 in steps of 0.1, 1-10 in steps of 1, etc., with ticks corresponding to 1, 10, and 100 drawn slightly longer.

## 5. PHASE LIFETIME

One of the common uses of phasor analysis is in estimating the lifetime of decays whose phasor is located on or in close proximity to the SEPL. For continuous PSEDs, Eq. (15) yields a simple relation between the phasor's phase $\varphi = \arg(z)$ and the lifetime $\tau$ on one hand, and between the modulus $m = |z|$ and the lifetime on the other [24]:

$$\begin{cases} \tau_\varphi = \dfrac{1}{2\pi f} \tan \varphi \\ \tau_m = \dfrac{1}{2\pi f m} \sqrt{1 - m^2} \end{cases} \tag{30}$$

These relations need to be modified for discrete phasors, based on the corresponding equations derived in the previous sections. For instance, based on the definition of the discrete phasor of PSEDs with a Dirac IRf (Eq. (18)), a phase and modulus lifetimes can be computed with:

$$\left. \begin{aligned} m \leq 1 &\Rightarrow \tau_m = \tau_m^- \\ 1 < m \leq \left|\cos\dfrac{\alpha}{2}\right|^{-1} &\Rightarrow \tau_m = \tau_m^+ \end{aligned} \right\} \rightarrow \begin{cases} \tau_\varphi = \dfrac{\theta}{\ln\left[\cos\alpha\left(1 + \dfrac{\tan\alpha}{\tan\varphi}\right)\right]} \\ \tau_m^\pm = \dfrac{\theta}{\ln\left(\dfrac{1-m^2}{\left(\sqrt{1-m^2\cos^2\dfrac{\alpha}{2}} \pm m\left|\sin\dfrac{\alpha}{2}\right|\right)^2}\right)} \end{cases} \tag{31}$$

where $\alpha$ is defined in Eq. (18).

As discussed in Section 3, the expression for the discrete phasor of gate PSEDs is in general more complex and therefore, no simple analytical relation between the phase or modulus and the lifetime can be derived, except in the few cases discussed in that Section, which reduce to the previous case up to a rotation. This does not mean that no relation exists between $\tau$ and $m$ or $\varphi$, but only that these relations are implicit only, and need to be computed numerically (using for instance the Phasor Explorer software).

## 6. PHASOR CALIBRATION

We have mentioned phasor calibration a number of times throughout this article. Due to space constraints, we will keep its discussion qualitative, and refer the reader to ref. [18] for an extended version.

As noted in Section 3, continuous phasor analysis is conceptually simple, as all details of the experiments intervene as constant (complex) multiplication factors in the final expression of the phasor. In other words, by dividing the final expression by the product of these constant factors (comprising the IRF's phasor, which, in its general sense, includes gating and/or binning), the final "normalized" phasor is identical to that defined in Eq. (15) for PSEDs and equal to the phasor of the "pure" decay (that is, the decay resulting from excitation and detection by a Dirac comb IRF) in case of other functions. The standard way this pre-factor or *calibration factor* is obtained (as the actual IRF might not necessarily be known), is by using a "calibration" PSED of known lifetime $\tau_C$, and obtaining the phasor of the IRF as:

$$z[I_T] = \frac{z\left[I_T \underset{T}{*} \Lambda_{\tau_C,T}\right]}{\zeta_f(\tau_C)} \quad (32)$$

In the previous equation, $z\left[I_T \underset{T}{*} \Lambda_{\tau_C,T}\right]$ is referred to as the *calibration phasor*, or phasor of the reference decay before calibration (a quantity that can be computed from the data) and $\zeta_f(\tau_C)$ is the *calibrated phasor* of the reference sample, given by the simple formula of Eq.(15).

Once such a calibration factor is obtained, any newly computed experimental (or uncalibrated) phasor $z[S_T]$ can be corrected (or calibrated) using it, according to:

$$\tilde{z}[S_T] \triangleq \frac{z[S_T]}{z[I_T]} = z[F_{0,T}] \quad (33)$$

where the notation of Eq. (13) are used and $F_{0,T}$ represents the decay emitted upon excitation by a Dirac comb ("pure" decay). Following this procedure, the calibrated phasor of any PSED ends up on the universal semicircle $L_\infty$ and the phasor of any other decay where it would be found, had the IRF been a Dirac comb.

This familiar result is not affected by decay offset, but decay truncation will in general break it down. This can be partially fixed by using an ad-hoc phasor harmonic frequency (see Section 4.2), but in general a naïve application of Eq. (32)-(33) will result in phasors that are close to those of the pure decays in the vicinity of the reference decay, but potentially departing from them away from the reference. Careful study of each non-ideal case is therefore recommended, in order to guide the choice of an appropriate reference decay (or change acquisition settings).

Unfortunately, such a simple procedure has an even more limited use case for discrete phasors. For one, as we have seen in Section 3, the SEPL of discrete decays is in general different from the universal circle. While there are simple cases (used as illustrations in Section 3.4) where the discrete phasor of a PSED is a simple product of a complex number with the discrete phasor of the "pure" PSED, $\zeta_{f,N}(\tau)$ (Eq. (18), this is the exception rather than the rule. In these favorable cases where the discrete phasors of PSEDs can be rewritten:

$$z_N\left[I_T \underset{T}{*} \Lambda_{\tau,T}\right] = \kappa z_N[I_T] z_N[\Lambda_{\tau,T}] \quad (34)$$

where $\kappa$ is a complex constant, then the calibration procedure described for continuous phasor applies. The main difference is that the calibration factor:

$$\frac{z_N\left[I_T \underset{T}{*} \Lambda_{\tau_C,T}\right]}{\zeta_{f,N}(\tau_C)} = \kappa z_N[I_T] \tag{35}$$

is now equal to the discrete phasor of the IRF times the $\kappa$ constant. Calibration of a computed phasor entails dividing it by the calibration factor given by Eq. (35), which is different from the phasor of the IRF. Moreover, the phasor of PSEDs are not brought back to the universal semicircle by this calibration procedure, but instead, to one of the SEPLs noted $\mathcal{L}_N$ in Section 3.4.

## 7. CONCLUSION

In this article, we have discussed some subtleties in the phasor analysis of periodic decays arising from the presence of a few non-idealities such as offset, truncation and most importantly, finite sampling and binning or time-gating. In particular, we have examined a few simple analytical examples of the modifications to the standard formulas generally used in phasor analysis, and pointed out some differences which it appears important to be aware of. The conclusion to draw from this work is certainly not that phasor analysis is of limited general use, but instead, that there exists a number of situations where careless application of formulas or procedures derived for the continuous case, may lead to erroneous results and conclusions. Importantly, binned data (e.g. TCSPC data) can be treated as discrete sampled data, for which the locus of phasor of single-exponential decays is an arc of circle, which converges to the universal semicircle as the number of sampling points increases.

## 8. ACKNOWLEDGEMENTS

This work was funded in part by Human Frontier Science Program Grant RGP0061/2015, US National Institute of Health Grants R01 GM095904 & R01 CA250636, University of California CRCC Grant CRR-18-523872, US Department of Energy Grant DE-SC0020338, the European Research Council under European Union's Horizon 2020 research and innovation program under grant agreement No. 669941 and by the Partner University Fund, a program of the French American Culture Exchange.